\documentclass[iop]{emulateapj}
\usepackage{apjfonts}
\usepackage{natbib}
\usepackage{graphicx}
\usepackage[usenames,dvips]{color}
\usepackage{url}

\def\Lx{L$_x$~}

\def\arcsec{$''$}
\newcommand{\Msun}{\ifmmode {M_{\odot}}\else${M_{\odot}}$\fi}
\newcommand{\Lsun}{\ifmmode {L_{\odot}}\else${L_{\odot}}$\fi}
\newcommand{\Rsun}{\ifmmode {R_{\odot}}\else${R_{\odot}}$\fi}

\newcommand{\refbf}[1]{#1}

\def\Chandra{${\it Chandra}$}

\def\HST{${\it HST}$}
\usepackage{xargs}
\usepackage{color}
\usepackage[normalem]{ulem} 

\shorttitle{A Low-Mass Main-Sequence Star and Accretion Disk in the Very Faint X-ray Transient M15 X-3}
\shortauthors{Arnason {\it et al.}}

\begin{document}
\title{A Low-Mass Main-Sequence Star and Accretion Disk in the Very Faint X-ray Transient M15 X-3}

\author{%
  R.~M. Arnason\altaffilmark{1},
  G.~R. Sivakoff\altaffilmark{1},
  C.~O. Heinke\altaffilmark{1}$^{,}$\altaffilmark{2},
  H.~N. Cohn\altaffilmark{3}, 
  P.~M. Lugger\altaffilmark{3}
}

\altaffiltext{1}{Department of Physics, University of Alberta, CCIS 4-181, Edmonton, AB T6G 2E1, Canada; rarnason@ualberta.ca}
\altaffiltext{2}{Max-Planck-Institute für Radioastronomie, Auf dem Hügel 69, 53121, Bonn, Germany}
\altaffiltext{3}{Astronomy Dept., Indiana University, 727 East 3rd St., Bloomington, IN 47405}

%\slugcomment{}

\begin{abstract}
We present near-simultaneous \Chandra/\HST\ observations of the very faint ($L_{x} < 10^{36}$ erg s$^{-1}$) X-ray transient source M15 X-3, as well as unpublished archival \Chandra\ observations of M15 X-3. The \Chandra\ observations constrain the luminosity of M15 X-3 to be $< 10^{34}$ erg s$^{-1}$ in all observed epochs. The X-ray spectrum shows evidence of curvature, and prefers a fit to a broken power-law with break energy $E_{\rm break} = 2.7^{+0.4}_{-0.6}$ keV, and power law indices of $\Gamma_{1} = 1.3^{+0.1}_{-0.2}$ and $\Gamma_{2} = 1.9^{+0.2}_{-0.2}$ over a single power law. We fit our new F438W ($B$), F606W (broad $V$), and F814W ($I$) {\it HST} data on the blue optical counterpart with a model for an accretion disk and a metal-poor main sequence star. From this fit, we determine the companion to be consistent with a main sequence star of mass $0.440^{+0.035}_{-0.060}$ $\Msun$ in a $\sim$4-hour orbit. X-ray irradiation \refbf{of the companion} is likely to be a factor in the optical emission from the system, which permits the companion to be smaller than calculated above, but larger than $0.15$ $M_{\odot}$ at the $3\sigma$ confidence level. M15 X-3 seems to be inconsistent with all suggested hypotheses explaining very faint transient behavior, except for magnetospherically inhibited accretion.
\end{abstract}


\maketitle
%----------------------------------------------------------------------

\section{Introduction}\label{s:intro}

	The Galactic globular clusters (GCs) are host to many exotic objects, due to the high central stellar density at their cores \citep{Clark75,VerbuntLewin06}. One such class of objects are the low-mass X-ray binaries (LMXBs), systems that consist of a compact object accreting material from a companion overflowing its Roche lobe. Many LMXBs exhibit transient behavior, where their luminosity can change by a factor of 100 or more. This behavior is generally believed to be tied to \refbf{thermal-viscous} instability in the accretion disk \citep{Lasota01}. 
	
	One broad category that is used to classify transient X-ray sources is their peak X-ray luminosity. Sources that have a peak luminosity exceeding $10^{37}$ erg s$^{-1}$ are classified as \textbf{\textit{bright}} sources. With the deployment of sensitive telescopes including \Chandra, \textit{XMM-Newton}, and \textit{Swift} in the past fifteen years, the class of \textbf{\textit{very faint}} LMXBs, whose peak X-ray luminosities lie below $10^{36}$ erg s$^{-1}$, have been discovered \citep{Muno05a,Wijnands06,Hands04,DegenaarWijnands09}. These objects present a puzzle for binary evolution, as their luminosity implies a time-averaged accretion rate that can reach as low as $10^{-13}$  $M_\odot {\rm \,  yr}^{-1}$ or less \citep{KingWijnands06}. Several explanations have been offered for these low accretion rates, including accretion from a planetary or brown dwarf companion, or by a primordial intermediate-mass black hole \citep{KingWijnands06}, accretion in an ultracompact system \citep{intZand05}, inefficient accretion due to magnetospheric inhibition \citep{Heinke09,DAngelo12,Heinke14}, or accretion from a companion in the period gap \citep{Maccarone13}. Some VFXTs can be explained as an intrinsically bright object viewed at an unfavourable inclination, but this is not possible for all such systems \citep{Muno05b,Wijnands06}. Observations of VFXTs in optical and infrared can constrain the nature of VFXT systems by identifying the companion in the system, so that the unexplained accretion regimes of these sources can be better understood. 

Recent X-ray monitoring campaigns of the Galactic Center have measured the long-term behavior of a number of VFXTs, identifying roughly as many objects that show only VFXT outbursts as those that show "normal", bright outbursts \citep{DegenaarWijnands09,DegenaarWijnands10,Degenaar12}.  Short, low-luminosity VFXT outbursts are also seen from transients showing bright outbursts \citep[e.g.,][]{DegenaarWijnands09}.  A few systems are seen to maintain quasi-persistent (for at least several continuous years) X-ray emission, well below $10^{36}$ erg s$^{-1}$, including NGC 6652 B \citep{Heinke01,Stacey12}, AX J1754.2-2754 \citep{Sakano02,Chelovekov07,Degenaar11}, 1RXS J171824.2-402934 \citep{intZand05,intZand09},  XMMU J174716.1-281048 \citep{DelSanto07}, M15 X-3 \citep{Heinke09}, and 1RXH J173524.4-354013 \citep{Degenaar10}. These systems are too faint to keep normal-sized (i.e., few hours orbital period) accretion disks irradiated, leading \cite{intZand05} to propose that the orbital separation must be extremely compact (orbital period of 7 minutes in one case); however, such a short orbital period would produce a very high mass transfer rate \citep[e.g.,][]{Deloye03}, leading to a contradiction. \refbf{Additionally, the measurement of strong H$\alpha$ emission lines in at least one source \citep{Degenaar10} shows a hydrogen-rich donor star, ruling out an ultracompact system in that case.}

The X-ray spectra of Galactic Center VFXTs have typically been well-described by a power-law of photon index ranging from $\sim$1 -- 3 \citep{DegenaarWijnands10,Degenaar12}.  However, the spectra have typically not been of high quality due to the high absorption, the small effective area of \textit{Swift/XRT}, or the short exposures in \Chandra\ or \textit{XMM-Newton} monitoring observations.  Three of the quasi-persistent VFXTs listed above have long \textit{XMM-Newton} exposures at luminosities of $10^{34}$ to $10^{35}$ erg s$^{-1}$, showing that all have rather soft spectra, with photon indices $\Gamma$=2.3-2.5, but that the objects fainter than $10^{35}$ erg s$^{-1}$ require a blackbody-like component in addition to a power-law \citep{ArmasPadilla13}. Another transient VFXT, IGR J17494-3030, observed with XMM-Newton at $L_X\sim8\times10^{34}$ erg s$^{-1}$, also requires a blackbody plus power-law in its spectral fit \citep{ArmasPadilla13b}.  

The high extinction towards most VFXTs has hampered deep searches for optical or infrared counterparts.  However, a few likely or certain optical/infrared counterparts have been identified.  \cite{Deutsch98} identified a relatively blue object as a likely LMXB counterpart in NGC 6652, which \cite{Heinke01} showed to be the counterpart to NGC 6652 B; this object is extremely variable in the optical and X-ray on minute timescales \citep{Engel12}.  
\cite{Degenaar10} found an optical counterpart for 1RXH J173524.4-354013, which shows a strong hydrogen emission line and magnitudes characteristic of a low-mass main sequence star. \cite{Kaur12} identified a likely variable near-infrared counterpart to XMMU J174716.1-281048, indicating a low-mass star.  The likely VFXT, and black hole candidate, Swift J1357.2-0933 has a low extinction, and a \refbf{potential M4 star companion \citep{Rau11}, although this potential companion was not detected when the system was quiescent \citep{Shahbaz13}. }

M15 X-3 was discovered from archival \Chandra\ High Resolution Camera (HRC-I) and High Energy Transmission Grating Spectrometer (HETGS) observations, by \cite{Heinke09}. This source, located roughly 21" from the center of the globular cluster M15, has been observed at an $L_X$ of either $\approx 10^{34}$ erg s$^{-1}$ (in its bright state) or $\approx 2 \times 10^{31}$ erg s$^{-1}$ (in quiescence). In zeroth-order HETGS observations M15 X-3 has demonstrated a relatively hard, power law-like spectrum ($\Gamma = 1.5$) in its bright state. It is likely to be a softer source in quiescence \citep{Heinke09}, though a spectrum of the faint state of M15 X-3 would require a long (> 50 ks) observation with \Chandra\, triggered when M15 X-3 is known to be in quiescence.  \cite{Heinke09} identified the optical counterpart to M15 X-3 as a likely main-sequence star, though the photometry had rather low signal-to-noise ratios. Throughout this analysis, we assume that M15 is at a distance of 10.3 kpc, with a column density of $4.6 \times 10^{20} {\rm \, cm^{-2}}$ \refbf{for consistency with \cite{Heinke09}}, an $E(\bv)$ of 0.10, and a cluster metallicity [Fe/H] of $-2.37$ \citep{Janulis92,vandenBosch06}.

\section{X-Ray Data Analysis}\label{s:X-ray}

\subsection{ACIS-S Data Reduction}

\begin{figure}
\includegraphics[scale=0.6]{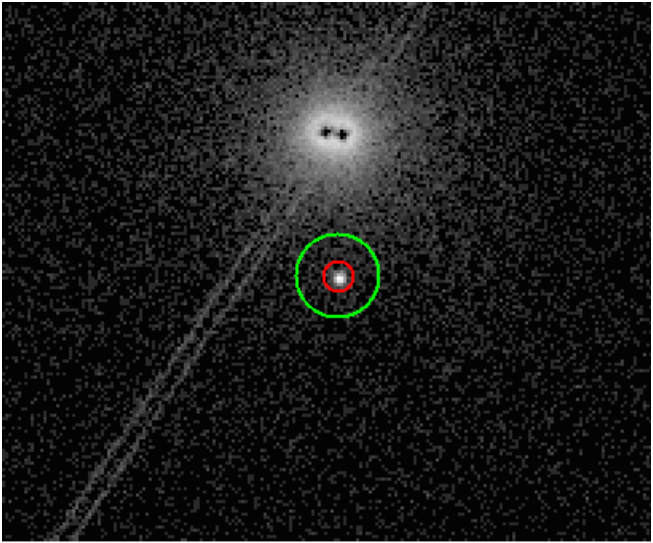}
\caption{X-ray image of M15 (ObsID 11029). The inner circle (radius 2.4") indicates the source extraction region for M15 X-3, and the inner boundary of the background extraction region. The outer circle (radius 6") is the outer boundary of the background extraction region. Note the presence of the heavily piled up LMXB sources AC-211 (left) and M15 X-2, as well as the \Chandra\ ACIS-S readout streak. M15 X-3 lies approximately 21" from the center of M15, well outside the $<$ 1.5" core \citep{SosinKing97}.	
\label{xrayimage}
}
\end{figure}

For this analysis, we reduced and analyzed the 2012 Advanced CCD Imaging Spectrometer (ACIS-S) \Chandra\ observation of M15, taken in subarray mode with an effective frame time of 0.9 s. We also reduced and analyzed three unpublished archival observations from the \Chandra\ data archive (see Table \ref{observationlog}). These observations were reduced using CALDB version 4.5.9 and the August 2012 time-dependent gain file (acisD2012-08-01t\_gainN0006.fits). All four observations were taken using ACIS-S in faint mode with no grating. Each observation was reprocessed with the chandra\_repro script, to generate a new level 2 events file. Using CIAO 4.6, the spectrum of M15 X-3 was obtained for each observation via the specextract script. In all observations, M15 X-3 can be clearly detected as a source 21" from the core of M15 (see Figure \ref{xrayimage}). Two other bright X-ray sources, the LMXBs AC-211 and M15 X-2 \citep{White01}, are also nearby in the \Chandra\ frame. The source region used was a 2.4" circular region, surrounded by an annulus with 6" radius that provided a representative background region. \refbf{Background regions of varying sizes were tested, however the choice of background region did not affect our results.} The spectra were grouped to give a minimum of 25 counts per bin in the 0.5-10 keV range, so that Gaussian statistics were valid.  Each spectrum was then fit using XSPEC 12.8.0. Initial fitting indicated that M15 X-3 was in a consistent spectral state across all four observations, motivating the choice to fit the spectra simultaneously, first with model parameters tied and then untied from each other. To perform a check for variability in M15 X-3, background subtracted light curves were generated using the dmextract command. Testing for variability using the CIAO tool glvary\footnote{http://cxc.harvard.edu/ciao/ahelp/glvary.html}, which looks for deviations among optimally sized bins \citep{GregoryLoredo92}, indicated evidence of variability within all 4 observations (probability of nonvariability always less than $10^{-8}$), as previously noted in other observations of M15 X-3 \citep{Heinke09}.

\subsection{HRC-I Data Reduction} 

There are five HRC-I observations of M15 X-3 in the \Chandra\ archive. The first three, obtained in 2001, found M15 X-3 in a quiescent state, barely detectable when using special processing \citep{Heinke09}, and we do not re-analyze them here. We analyze the 2007 and 2011 HRC-I observations, which catch M15 X-3 in a relatively bright state (see details in Table \ref{observationlog}). We reprocessed the events files (with CALDB 4.5.9 and the August 2012 time-dependent gainfile) using the chandra\_repro script, and extracted source counts from the position of M15 X-3, and background counts from an annular region. We infer the unabsorbed 0.5-10 keV luminosities using PIMMS and a power law with photon index $\Gamma = 1.4$  and $N_{H}$ = $4.6\times10^{20}$ cm$^{-2}$. The short length of these observations (Table \ref{observationlog}) makes the data unsuitable for timing analyses. 

%Figure: Image of the ACIS-S field, including identification of AC-211, M15 X-2, and M15 X-3

\begin{center}
\begin{deluxetable}{ccccc} 
  \tablecolumns{5}
  \tablewidth{0pt}
  \tablecaption{%
    Summary of \Chandra\  Observations of M15 X-3%
    \label{observationlog} 
  }
  \tablehead{%
   \colhead{ObsID} &
   \colhead{Date} &
   \colhead{Exposure} &
   \colhead{Instrument} &
   \colhead{$L_{x}$} \\
   \colhead{} & 
   \colhead{} &     
   \colhead{[ks]} &
   \colhead{} &
   \colhead{[$10^{33}$ erg s$^{-1}$]} 
    }
  \startdata
    \dataset[ADS/Sa.CXO#obs/09584]{\phn9584} & 2007-09-05 & \phn2.15 & HRC-I & $6^{+2}_{-2}$\\ 
    \dataset[ADS/Sa.CXO#obs/11029]{\phn11029} & 2009-08-26 & 34.18 & ACIS-S & $8.3^{+0.4}_{-0.4}$ \\ 
    \dataset[ADS/Sa.CXO#obs/11886]{\phn11886} & 2009-08-28 & 13.62 & ACIS-S & $9.9^{+0.7}_{-0.7}$ \\ 
    \dataset[ADS/Sa.CXO#obs/11030]{\phn11030} & 2009-09-23 & 49.22 & ACIS-S & $9.2^{+0.4}_{-0.4}$  \\ 
    \dataset[ADS/Sa.CXO#obs/13420]{\phn13420} & 2011-05-30 & \phn1.45 & HRC-I & $5^{+2}_{-1}$ \\ 
    \dataset[ADS/Sa.CXO#obs/13710]{\phn13710} & 2012-09-18 & \phn4.88 & ACIS-S & $10^{+1}_{-1}$ 
     \enddata
\end{deluxetable}
\end{center}

\subsection{M15 X-3 Spectral Analysis}

	Previously, M15 X-3 was found to be well-described using a power law with photon index $1.51 \pm 0.14$, absorbed through a column density fixed at the cluster value of $4.6 \times 10^{20} {\rm \, cm^{-2}}$ \citep{Heinke09}. Very faint X-ray transients are often described with a hard power-law, or a hard power-law with a softer thermal component \citep{ArmasPadilla13}. Even for a relatively faint source such as M15 X-3, the count rate of $\approx$ 0.06 counts s$^{-1}$ (or 0.18 per frame, for the three long archival ACIS-S observations), makes pileup a potential issue. Therefore, pileup is accounted for by convolving spectral models with the XSPEC model for pileup in \Chandra\ observations \citep{Davis01}. The dependence of fits upon the grade correction parameter $\alpha$ was considered in the fitting process \footnote{$N^{\alpha}$ gives the probability that the photon will be identified as a 'bad' grade and rejected, where $N$ is the number of photons landing in the detection cell during one frame time \citep{Davis01}.}. 
	
	For this analysis we fit all four observations simultaneously. These separate fits all suggested that M15 X-3 was in the same state with \Lx $\approx 8\times10^{33}$ erg s$^{-1}$ ($\chi^2 = 211.73$ for $188$ degrees of freedom). This motivated the choice to fit all four ACIS observations of M15 X-3 simultaneously, with the power-law index of all four observations tied together. In this case, only the power law component normalization, which corresponded to the measured flux of the source, was allowed to be independent. Fitting to all of the four \Chandra\ observations gives an acceptable fit ($\chi^2$/dof $= 214.13/191$, implying a null hypothesis probability of 0.12, see Table \ref{xrayfits} SPL 1) with an absorbed power law of photon index $1.42 \pm 0.03$. Allowing the indices to vary does not significantly improve the fit (an F-test gives an F-statistic of 0.71, and probability 0.55 for obtaining such an improvement by chance), suggesting that the power-law index does not change. The fit was not significantly affected by the value of the grade correction parameter $\alpha$, which was unconstrained in the fit, due to the relatively small degree of pileup. We therefore report results with $\alpha$ fixed to 0.5, as empirically suggested \citep{Davis01}, as none of the results presented here qualitatively change for different choices of $\alpha$.
	
\begin{figure}
\includegraphics[scale=0.35]{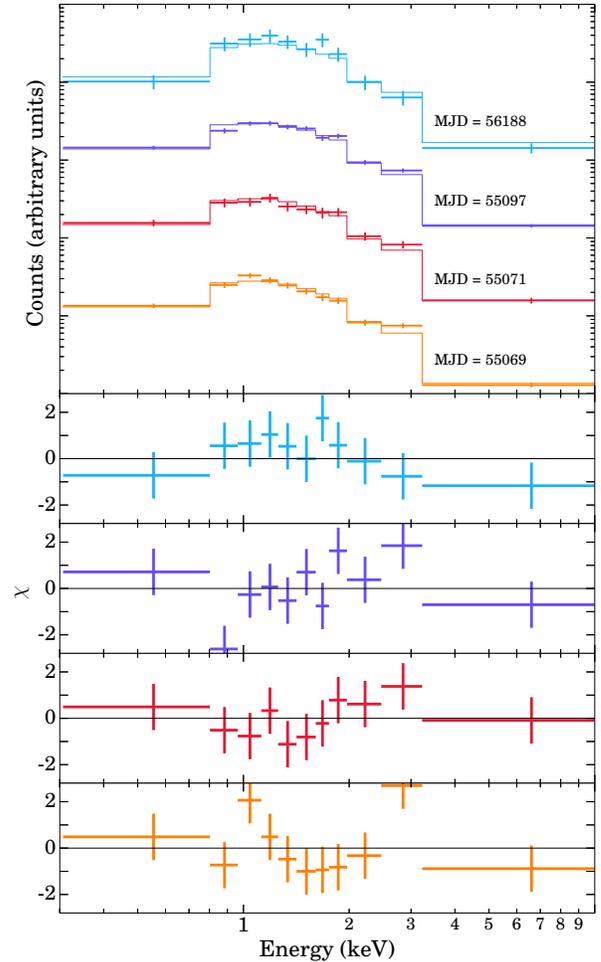}
\caption{%
X-ray spectra of M15 taken from the four ACIS-S \Chandra\ observations, fitted to a single power-law with column density as a free parameter. The spectra are sorted with the most recent at the top. \refbf{Note that many of the residuals lie above the model between 1 - 3 keV and all of the residuals lie below the model above 3 keV, indicating the potential presence of curvature in the spectrum (particularly for the MJD = 56188 epoch).} These data have been rebinned for clarity.
\label{spectrum}
}
\end{figure}

\refbf{Examining the single power-law fit suggests the presence of curvature in the residuals above 1 keV. This curvature motivated the choice to attempt other fits. First, the column density $N_{H}$ was allowed to vary, since high column density is a potential source of curvature. All subsequent fits permit $N_{H}$ to vary. In the case of the single power-law, this strongly improved the fit ($\chi^2$/dof $= 195.08/190$; SPL 2 in Table \ref{xrayfits}) with a new column density of $9^{+2}_{-1} \times 10^{20} {\rm \, cm^{-2}}$, though the residuals of the fit (Figure \ref{spectrum}) still suggested curvature. } Since other VFXTs are described acceptably by a spectra composed of a power-law with a thermal component \citep{ArmasPadilla13}, several thermal components (a low magnetic field neutron star hydrogen atmosphere, NSATMOS, \cite{Heinke09}; a multicolor disk blackbody, DISKBB, e.g. \cite{Mitsuda84}; and a \refbf{single-temperature} blackbody, BBODYRAD) were tested to see if they improved the fit. 

\refbf{Adding an NSATMOS component (assuming emission from the entire neutron surface, i.e., normalization fixed to 1.0; NSATMOS + PL in Table \ref{xrayfits}) does not substantially improve the fit ($\Delta\chi^{2}$ is only 0.04 for 1 fewer d.o.f), and the fit prefers a temperature of $0.05^{+0.03}_{-0.04}$ keV. The 90\% confidence upper limit on the temperature permits this component to contribute only 2\% ($\sim2\times10^{32}$ erg s$^{-1}$) of the 0.5---10 keV unabsorbed $L_X$, and the lower limit is pegged to the lower temperature boundary of the model. Adding a disk blackbody component (DISKBB + PL in Table \ref{xrayfits}) does improve the fit ($\chi^{2}$/dof $= 188.04/188$, F-statistic of 3.52 and probability $3.2\times10^{-2}$ of this improvement being due to chance). However, the disk blackbody model can be physically excluded, since the inferred inner disk radius ($R_{\rm inner} \cos \theta = $ 0.7 km) is much smaller than the neutron star radius, making an accretion disk origin untenable except at very high inclination.} 

\refbf{Addition of a blackbody model can also improve the fit ($\chi^{2}/dof = 187.52/188$, F-statistic of 3.79 and probability $2.4 \times 10^{-2}$ of this improvement compared to the variable $N_H$ single power-law model being due to chance). The best-fit blackbody parameters are $kT$=$0.7^{+0.06}_{-0.07}$ keV, with an emitting radius $R$=$0.2^{+0.1}_{-0.1}$ km (see Table \ref{xrayfits}, BB + PL).} This could possibly be interpreted as a hot spot on the neutron star surface, where infalling material is channeled by the magnetic field to the poles. However, the inferred emitting region is smaller than is expected for polar caps on a neutron star surface spinning at millisecond periods (1.4--2.6 km for 3-10 ms periods, using $R_{\rm pc}=(2 \pi R_{\rm NS}/(cP))^{1/2} R_{\rm NS}$; \citealt{LyneGrahamSmith06} ).  
A longer spin period (e.g., 0.5 seconds for 0.2 km) would predict a polar cap of the appropriate size.  

Alternatively, it is well-known (e.g. \citealt{Zavlin96}; \citealt{Rajagopal96}) that a hydrogen atmosphere model gives inferred radii a few times larger than blackbody fits to the same data.  Indeed, blackbody fits to the thermal spectra of old millisecond pulsars find inferred polar cap radii of order 0.1-0.2 km \citep{Bogdanov06,Forestell14}.  Using hydrogen atmosphere models, emitting radii of $\sim$1 km, more typical of NS polar caps, are found (e.g. \citealt{Zavlin02}; \citealt{BeckerAschenbach02}; \citealt{Bogdanov06}). We therefore try fitting the thermal component with the NSATMOS model again, now permitting radiation from only a small fraction of the stellar surface, by freeing the normalization parameter. \refbf{This result of this fit gave an aphysical normalization (corresponding to the fraction of the neutron star emitting) $\gg1$. Thus, so fitting was repeated with an alternate neutron star atmosphere NSA model \citep{Zavlin96}, with magnetic field set to 0 G, which is a reasonable approximation for the $10^{8-9}$ G regime (NSA + PL in Table \ref{xrayfits}).  The resulting fit is again an improvement over the single power-law fit ($\chi^{2}$/dof $= 187.99/188$, F-statistic of 3.55 and probability $3.1\times10^{-2}$ of this improvement compared to the variable $N_H$ single power-law model being due to chance).  The best-fit temperature is $0.6^{+0.3}_{-0.2}$ keV and the inferred radius of the emitting region is $0.3^{+0.4}_{-0.2}$ km, with the temperature upper error limit pegged to the upper boundary of the model.} 

Finally, we tried a broken power-law model, where the spectrum is described by a power law of index \refbf{$\Gamma_{1} = 1.3^{+0.1}_{-0.2}$ up to an energy of $2.7^{+0.4}_{-0.6}$ keV, where it becomes a power law of index $\Gamma_{2} = 1.9^{+0.2}_{-0.2}$. This model also gives a significantly better fit than the single power-law ($\Delta \chi ^{2}$ = 9.89 over the single power-law, F-statistic 5.02, probability 7.5$\times10^{-3}$ of this improvement compared to the variable $N_H$ single power-law model being due to chance)}. Releasing the power-law indices or break energies between observations did not improve this fit. The broken power-law model would presumably imply a bremsstrahlung origin of the emission, with the break energy suggesting the bremsstrahlung temperature \citep{Chakrabarty14}, so it is reasonable to search for a separate emission contribution from the neutron star surface in this case. However, as above, adding a NSATMOS component (with emitting size fixed to the whole neutron star) gave no improvement, and adding a BBODYRAD component gave very little improvement ($\Delta\chi^2$ = 0.01). \refbf{Although it is an empirical fit only, we adopt the broken power-law model as our best-fit as it gave the best improvement compared to the variable $N_H$ single power-law model.}

%Figure: M15 X-3 spectrum

\begin{deluxetable*}{ll|ccccccc}
\tablecolumns{9}
\tablewidth{0pt}
\tablecaption{  Summary of fits to \Chandra\ observations of M15 X-3 \label{xrayfits} }
\tablehead{%
 \colhead{Model} & & \colhead{SPL 1} & \colhead{SPL 2} & \colhead{BB + PL} & \colhead{DISKBB + PL} & \colhead{NSATMOS + PL } & \colhead{NSA + PL} & \colhead{BPL$\dagger$} }
\startdata
$N_{H}$ & [$10^{20}$ cm$^{-2}$] & $[4.6]$ & $9^{+2}_{-1}$ & $6^{+4}_{-4}$ & $4^{+5}_{-4}$ & $10^{+2}_{-2}$ & $5^{+5}_{-5}$ & $6^{+2}_{-3}$  \\
$\Gamma_{1}$ & & $1.42^{+0.03}_{-0.03}$ & $1.57^{+0.07}_{-0.07}$ & $1.5^{+0.2}_{-0.2}$ & $1.3^{+0.5}_{-0.5}$ & $1.57^{+0.04}_{-0.07}$ & $1.5^{+0.3}_{-0.3}$ & $1.3^{+0.1}_{-0.2}$  \\
$\Gamma_{2}$ & & \nodata & \nodata & \nodata & \nodata & \nodata & \nodata & $1.9^{+0.2}_{-0.2}$  \\
$E_{\rm break}$ & [keV] &\nodata & \nodata & \nodata & \nodata & \nodata & \nodata & $2.7^{+0.4}_{-0.6}$  \\
$kT_{\rm thermal}$ & [keV] & \nodata & \nodata & $0.7^{+0.3}_{-0.2}$ & $1.2^{+1.0}_{-0.3}$ & $0.05^{+0.03}_{-0.04*}$ & $0.6^{+0.3*}_{-0.2}$ & \nodata  \\
$R_{\rm thermal}$ & [km] & \nodata & \nodata & $0.2^{+0.1}_{-0.1}$ & $0.7^{+0.3}_{-0.5}$ & $[10]$ & $0.3^{+0.4}_{-0.2}$ & \nodata  \\
$L_{X}$, MJD = 55069 & [$10^{34}$ \rm erg s$^{-1}$] & $0.83^{+0.04}_{-0.04}$ & $0.83^{+0.04}_{-0.04}$ & $0.77^{+0.05}_{-0.05}$ & $0.74^{+0.06}_{-0.06}$ & $0.83^{+0.04}_{-0.04}$ & $0.76^{+0.05}_{-0.05}$ & $0.75^{+0.05}_{-0.05}$  \\
PL fraction, MJD = 55069 & & [1] & [1] & $0.88^{+0.07}_{-0.09}$ & $0.7^{+0.2}_{-0.3}$ & $0.99^{+0.01}_{-0.01}$ & $0.84^{+0.04}_{-0.05}$ & [1]  \\
$L_{X}$, MJD = 55071 & [$10^{34}$ \rm erg s$^{-1}$] & $0.99^{+0.07}_{-0.07}$ & $0.98^{+0.07}_{-0.07}$ & $0.93^{+0.08}_{-0.08}$ & $0.91^{+0.09}_{-0.09}$ & $0.98^{+0.07}_{-0.07}$ & $0.92^{+0.08}_{-0.08}$ & $0.90^{+0.08}_{-0.07}$  \\
PL fraction, MJD = 55071 & & [1] & [1] & $0.90^{+0.06}_{-0.08}$ & $0.8^{+0.2}_{-0.2}$ & $0.99^{+0.01}_{-0.01}$ & $0.87^{+0.03}_{-0.04}$ & [1] \\
$L_{X}$, MJD = 55097 & [$10^{34}$ \rm erg s$^{-1}$] & $0.92^{+0.04}_{-0.04}$ & $0.92^{+0.04}_{-0.04}$ & $0.87^{+0.05}_{-0.05}$ & $0.85^{+0.06}_{-0.07}$ & $0.92^{+0.04}_{-0.04}$ & $0.86^{+0.06}_{-0.06}$ & $0.84^{+0.06}_{-0.03}$  \\
PL fraction, MJD = 55097 & & [1] & [1] & $0.89^{+0.07}_{-0.09}$ & $0.7^{+0.2}_{-0.2}$ & $0.99^{+0.01}_{-0.01}$ & $0.86^{+0.03}_{-0.03}$ & [1]  \\
$L_{X}$, MJD = 56188 & [$10^{34}$ \rm erg s$^{-1}$] & $1.0^{+0.1}_{-0.1}$ & $1.0^{+0.1}_{-0.1}$ & $1.0^{+0.1}_{-0.1}$ & $1.0^{+0.1}_{-0.1}$ & $1.1^{+0.1}_{-0.1}$ & $1.0^{+0.1}_{-0.1}$ & $1.0^{+0.1}_{-0.1}$  \\
PL fraction, MJD = 56188 & & [1] & [1] & $0.91^{+0.06}_{-0.08}$ & $0.8^{+0.2}_{-0.2}$ & $0.99^{+0.01}_{-0.01}$ & $0.88^{+0.03}_{-0.04}$ & [1] \\
$\chi^{2}/d.o.f$ & & $214.13/191$ & $195.08/190$ & $187.52/188$ & $188.04/188$ & $195.03/189$ & $187.99/188$ & $185.19/188$  \\
n.h.p. & & $0.12$ & $0.38$ & $0.50$ & $0.49$ & $0.37$ & $0.49$ & $0.54$  

\enddata
\tablecomments{%
    \refbf{M15 X-3 best spectral fit parameters, all ACIS-S data. Values in square brackets are fixed as part of the model choice. Error values marked with $*$ indicate a model limit. In this table SPL 1 is the single power-law with column density fixed, while SPL 2 is the same but with column density free. $\dagger$ We adopt the broken power-law model as our best-fit.}
    }
\end{deluxetable*}

\section{Optical Data Analysis}
 
%Figure: Finding chart for M15 X-3 optical counterpart   
\begin{figure}
\includegraphics[scale=0.5]{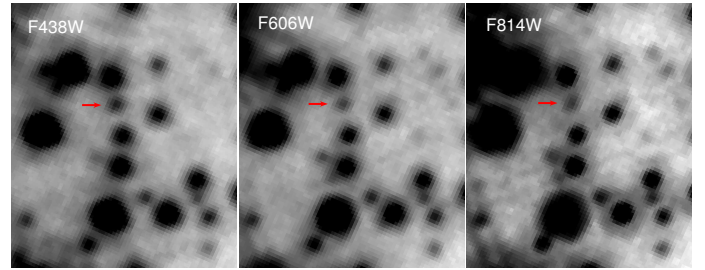} 
\caption{Finding charts for the optical counterpart of M15 X-3. N is up and E is to the left. The field of each frame is 1.4\arcsec\ wide by 1.6\arcsec\ tall. In each frame, the location of M15 X-3 is identified with a red arrow. All images are drizzle-combined frames from the HST/WFC3 observation taken in 2012 (MJD=56187), nearly simultaneous with Chandra ObsID 13710.
\label{findingchart}
} 
\end{figure}

%Figure: CMD for M15, including identification of M15 X-3's counterpart
\begin{figure}
\includegraphics[scale=0.4]{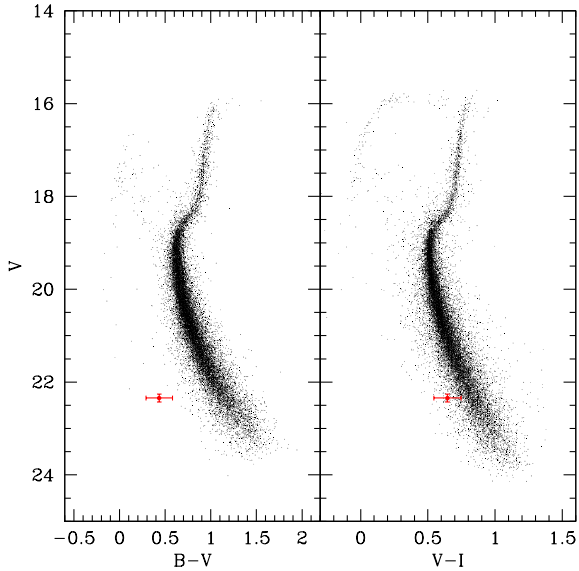} 
\caption{Color-magnitude diagram for M15, from 2012 \HST/WFC3 data. The red dot indicates the location of the optical counterpart for M15 X-3, with errors calculated based on the half-width of the main sequence at the magnitude of the counterpart. Note that M15 X-3 appears to be bluer than the main sequence, especially as one moves to bluer filters. Aside from M15 X-3's optical counterpart, only stars with small formal photometric error ($< 0.025$ mag) were kept to clean the image.
\label{CMD}
} 
\end{figure}  

\begin{deluxetable}{cccccc}
 \tablecolumns{6}
 \tablewidth{0pt}
 \tablecaption{%
  Summary of \HST\ Observations of M15 X-3%
  \label{HSTtable}
  }
  \tablehead{%
   \colhead{Proposal ID} &
   \colhead{Date} & 
   \colhead{MJD} &
   \colhead{Filter} & 
   \colhead{Instrument} & 
   \colhead{Magnitude} 
  }
  \startdata
   5742 & 1994-10-26 & 49651 & F336W & WFPC2 & $21.5 \pm 0.2$ \\
   9039 & 2002-04-05 & 52369 & F555W & WFPC2 & $22.0 \pm 0.2$ \\
   9039 & 2002-04-05 & 52369 & F439W & WFPC2 & $23.7 \pm 0.8$ \\
   12751 & 2012-09-17 & 56187 & F814W & WFC3 & $21.69 \pm 0.04$ \\ 
   12751 & 2012-09-17 & 56187 & F606W & WFC3 & $22.34 \pm 0.09$ \\
   12751 & 2012-09-17 & 56187 & F438W & WFC3 & $22.77 \pm 0.12$
   \enddata
   \tablecomments{ Observations in 2012 were calibrated to the VEGAMAG system, while observations in 1994 and 2002 were calibrated to the STMAG system, as per \cite{Heinke09} }
  \end{deluxetable} 

\subsection{WFC3 Photometery}

The 2012 \HST\ observation, taken in one orbit with the WFC3 camera using the UVIS channel in subarray mode, one day before a \Chandra\ observation, consists of exposures in WFC3 $B$ (F438W, total exposure length 1360s), broad $V$ (F606W, total exposure length 188s) and $I$ (F814W, total exposure length 332s). The exposures were obtained using the standard 4-point box dither pattern, with integer plus subpixel dithers, to allow for bad pixels and increase the effective angular resolution. A 1k$\times$1k subarray was chosen to minimize the overhead time and thus allow the necessary exposures to fit into a single orbit. In programming the observations, care was taken to constrain the roll angle in order to prevent diffraction spikes from a nearby bright giant from falling on the M15 X-3 counterpart.  The frames were combined with correction for cosmic rays using the IRAF/STSDAS Multidrizzle task with $2\times$ oversampling, giving an effective pixel size of 0.02\arcsec. Photometry on the drizzle-combined frames was carried out using the DAOPHOT/ALLSTAR software package, since the field of M15 rapidly becomes crowded near the core. PSFs were constructed using 100 candidate PSF stars per filter. Photometric calibration to the VEGAMAG system was performed by doing aperture photometry on the PSF stars within a 0.1\arcsec\ diameter aperture, finding the aperture correction to an infinite radius aperture from \citet{Sirianni05}, calculating the offset between the ALLSTAR photometry and the aperture photometry, and applying the calibrations from the HST WFC3 calibration webpage \footnote{http://www.stsci.edu/hst/wfc3/phot\_zp\_lbn}. 

As shown in Figure \ref{findingchart}, the optical counterpart of M15 X-3 is clearly detected in all 3 filters. Its magnitudes in this epoch, as well as in previous usable \HST\ epochs, are given in Table \ref{HSTtable}. CMDs of the cluster were constructed in $B$ -- $V$ and $V$ -- $I$, as shown in Figure \ref{CMD}, where the counterpart is bluer than the main sequence and becomes progressively bluer as one moves to bluer filters. 

\subsection{Pysynphot Modelling}
To determine the nature of the M15 X-3 companion using its optical colors, the system is modelled using Pysynphot \citep{Pysynphot}. Previous observations of M15 X-3 in the X-ray and optical bands have suggested that the system is likely to be a neutron star with a low-mass main sequence companion \citep{Heinke09}. M15 X-3 appears to be too X-ray luminous to be a WD accretor \citep{Verbunt97} and its X-ray/optical flux ratio rules this out conclusively \citep{Heinke09}. The optical spectrum of the M15 X-3 system is therefore modelled with two components: a low-mass main sequence companion and an accretion disk whose emission is approximated by a power law. 

\refbf{The range of power-law indices used to model a power-law accretion disk (and its expected optical spectrum) depends on two limiting behaviours.} For small frequencies, $F_{\nu}$ approaches a Rayleigh-Jeans description where $F_{\nu} \propto \nu^{2}$. For a conventional disk, the spectrum of the outer part of the disk behaves approximately as $F_{\nu} \propto {\nu}^{1/3}$ \citep{FKR02}. In the wavelength space utilized by Pysynphot for spectral modelling, this corresponds to a photon index $\Gamma_{\lambda} = -4$ in the Rayleigh-Jeans limit, and $\Gamma_{\lambda} = -7/3$ for a conventional disk. In fitting the power law, $\Gamma_{\lambda}$ indices between $-4$ and $-2$ were iconsidered. To have an index larger than $-2$ (which corresponds to the peak of the spectrum) would imply that the measured WFC3 bandpasses lie close to the Wien limit of the spectrum. Observations of quiescent disks in other accreting NS and black hole systems \citep[e.g.][]{Froning11,Hynes12,Wang13} indicate that the disks have typical inner temperatures between 5000 and 13000 K. \refbf{Compared to Aql X-1 in quiescence, M15 X-3 is more luminous in both the X-rays (by a factor of ~20;  \citealt{Narayan97}) and UV (by a factor of $\sim$4; \citealt{Hynes12}). Assuming they have a similar temperature-radius relation, we conclude that M15 X-3 is likely hotter than Aql X-1. Since Aql X-1's disk temperature was $12400 \pm 1400$ K \citep{Hynes12}, it is unlikely that the disk temperature of M15 X-3 is low enough to place the Wien tail into our WFC3 bandpasses.}

%As M15 X-3 is in a higher accretion state than true quiescence, its disk should be hotter, making disk temperatures low enough to place the Wien tail into our WFC3 bandpasses unlikely.  
%We note that if the accretion disk of M15 X-3 extended to the Alfvén radius, then the inner temperature would be roughly 40,000 to 100,000 K for NS magnetic fields in the range $10^{9} - 10^{8}$ G, consistent with our bandpasses occuring well redward of the peak of the spectrum \cite{FKR02}. 

The power-law component is processed with Synphot's built in power-law model, while the low-mass companion is modelled using stellar models from the Castelli and Kurucz library of stellar atmospheres included in the Synphot package. The advantage of this library is that it permits the modelling of main-sequence stars at the low metallicity characteristic of M15. As inputs, the Castelli and Kurucz models require the following: effective temperature $T_{\rm eff}$, logarithm of surface gravity $ \log(g)$, and metallicity $[M/H]$. To obtain these input parameters \refbf{for a main-sequence star of a particular mass}, results were taken from \cite{Baraffe97} model \refbf{calculations} for low-mass stars at $[M/H] = -2.0$, which is reasonable given that M15 has $[{\rm Fe/H}] = -2.34$ \citep{Carretta09}. \refbf{The nature of the companion and disk is determined by fitting the B, V, and I magnitudes predicted by (Synphot) synthetic photometry for a particular choice of low-mass companion and accretion disk against the measured WFC3 magnitudes for the M15 X-3 system using a standard chi-square test. A grid-search for the best chi-square value is performed over the parameter space for each free component --- mass in the case of the companion, and power-law index in the case of the accretion disk. Four cases were considered: A low-mass main-sequence companion only, a power law accretion disk only (with free index), a companion and power-law disk with a free index, and a companion and power-law disk with index fixed to $-7/3$. In the case of a low-mass companion + free index power-law disk,} testing only at the mass values given in \cite{Baraffe97} gives a best-fit companion mass of $0.5 \pm 0.1 M_{\odot}$, with a power-law index $\Gamma_{\lambda} = -3.28$. To estimate a more precise mass, linear interpolation of model parameters was performed on the table of masses in the range $0.200-0.700 \, M_{\odot}$. 

\begin{figure}
\includegraphics[scale=0.4]{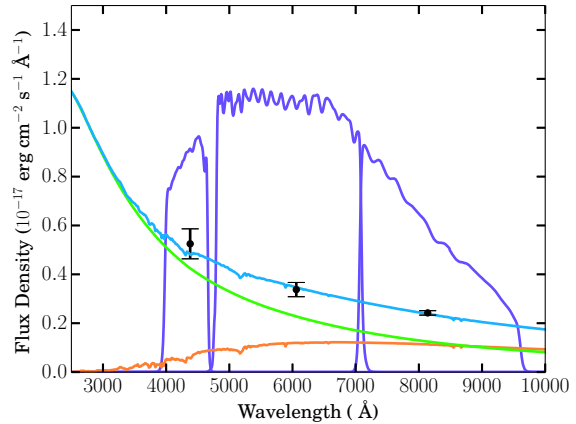}
\caption{Inferred optical spectrum for M15 X-3. The spectrum of the $0.440$ $M_{\odot}$ companion is plotted in orange. The spectrum of the accretion disk, modelled as a power-law, is plotted in green. The combined spectrum is plotted in cyan, with the black points representing measured optical fluxes in each of the three \HST\ filters. The bandpasses of each of the three \HST\ filters are plotted in arbitrary units in purple for reference. 
\label{inferredspectrum}
}
\end{figure}

\begin{deluxetable*}{lcccccccccc} 
  \tablecolumns{10}
  \tablewidth{0pt}
  \tablecaption{ Summary of fits to \HST\ observations of M15 X-3   \label{opticalfit} }
  \tablehead{%
   \colhead{Model} &
   \colhead{$\Gamma_{\lambda}$} &
   \multicolumn{6}{c}{$F_{\rm disk}/F_{\rm total}$} & 
   \colhead{Inferred $M_{\rm comp}$} & 
   \colhead{$\chi^{2}/dof$} \\  
   \colhead{} &
   \colhead{} & 
   \colhead{$U_{1994}$} &
   \colhead{$B_{2002}$} &
   \colhead{$V_{2002}$} &
   \colhead{$B_{2012}$} &
   \colhead{$V_{2012}$} &
   \colhead{$I_{2012}$} &
   \colhead{} &
   \colhead{} &
    }
  \startdata
Companion only & \nodata & [0] & [0] & [0] & [0] & [0] & [0] & $0.540 \pm 0.005$ & 33.78/2 \\ 
PL only & $-2.00^{+0.00*}_{-0.06}$ & [1] & [1] & [1] & [1] & [1] & [1] & \nodata & 8.87/1 \\ 
Fixed PL + companion  & $[-\frac{7}{3}]$ & $0.98^{+0.01}_{-0.01}$  & $0.92^{+0.04}_{-0.04}$ & $0.82^{+0.07}_{-0.07}$ &  $0.87^{+0.07}_{-0.06}$ & ${0.68}^{+0.13}_{-0.13}$ & ${0.53}^{+0.15}_{-0.15}$ & $0.440^{+0.035}_{-0.060}$ & 0.44/1 \\
Free PL + companion & $-4.00^{+2.00*}_{-0.00*}$ & $0.94^{+0.05}_{-0.01}$ & $0.85^{+0.14}_{-0.03}$  & $0.62^{+0.34}_{-0.02}$ & $0.71^{+0.28}_{-0.07}$ & $0.35^{+0.58}_{-0.03}$ & $0.16^{+0.71}_{-0.01}$ &  $0.515^{+0.005}_{-0.270}$ & 0.18/0 
\enddata
\tablecomments{%
    Summary of fits to 2012 \HST\ data of M15 X-3 optical counterpart using PYSYNPHOT modelling. Values in square brackets were fixed because of the model choice, while uncertainties marked with a single asterisk indicate that the uncertainty is at the hard limit of the model. \refbf{Note that adding a power-law with free index to the model adds two model parameters (PL normalization, PL index), while a power law with a fixed index adds only one model parameter (PL normalization). A low-mass main-sequence companion adds only one free parameter to the model (mass of the companion).}
  }

\end{deluxetable*}

As shown in Table \ref{opticalfit}, neither a power-law disk description nor a companion alone is sufficient to explain the observed optical colors of M15 X-3. The best fit is obtained by having a combined spectrum, where the accretion disk of M15 X-3 and its low-mass companion both contribute to the optical flux. Allowing the mass of the companion as well as both power law parameters ($\Gamma_{\lambda}$ and the normalization) to vary gives an acceptable fit with an inferred companion mass of $0.515^{+0.005}_{-0.270}$ \Msun. However, with this many free parameters, the degrees of freedom reduce to zero, overconstraining the model and limiting its utility. \refbf{Such a fit is only useful for constraining the model parameters --- in this case, the fit shows the stellar component is not consistent with a low stellar mass.} If the index is fixed at the conventional accretion disk value of $-7/3$ an acceptable fit is also obtained, with an inferred companion mass of $0.440^{+0.035}_{-0.060}$ \Msun. \refbf{The inferred spectrum of this combined system is plotted in Figure \ref{inferredspectrum}.} This choice of index describes the accretion disk as the largest contributor to the system's optical flux in B and V, while in I the flux contribution of the disk and companion become approximately equal. This naturally explains why M15 X-3's optical counterpart appears much bluer than the main sequence, especially in bluer filters, as it is the accretion disk which dominates emission at bluer wavelengths. \refbf{This fit has $\chi^2=0.44$ for 1 degree of freedom. With this small d.o.f., this low $\chi^2$ corresponds to a null hypothesis probability of 0.49 (close to the null hypothesis probability of 0.5 for a reduced $\chi^2 = 1$ for large d.o.f.), which indicates that there is no need to consider additional systematic errors when constraining our fit.} Note that the fits prefer a power-law photon index of $-4$, consistent with a relatively high temperature for the accretion disk.

\subsection{X-ray Irradiation of the Companion}

Given the relative blueness of the optical counterpart of M15 X-3 compared to other members of the cluster, it is reasonable to ask whether its optical colors may be substantially altered by X-ray irradiation of the main sequence companion of M15-3 by the primary. To determine whether X-ray irradiation is important, we compare the amount of X-ray light received by the companion star (measured near-simultaneously) with the total flux from the star. We begin by assuming a non-irradiated companion, $M_{2} = 0.44 \, M_{\sun}$ and a $1.4 \, M_{\sun}$ primary that radiates $\approx 10^{34}$ erg s$^{-1}$ of X-rays. With a mass ratio $q = M_{2}/M_{1}$ one can write the binary separation as (following \citealt{FKR02}):

\begin{equation}
a = 3.5 \times 10^{10} (M_{1})^{\frac{1}{3}} ( 1 + q )^{\frac{1}{3}} (P_{\rm hr})^{\frac{2}{3}} \text{cm}.
\end{equation}

For a lower MS star filling its Roche Lobe, the period can be approximated by
\begin{equation}
M_{2} \approx 0.11 P_{\rm hr} ,
 \end{equation}
 
\noindent which means that the orbital period of the system is approximately 4 hours \citep{FKR02}. Substituting this relation in for $M_{2}$ gives the following:

\begin{equation} 
a \simeq 3.5 \times 10^{10} M_{1} ( 1 + q )^{\frac{1}{3}} \left(\frac{M_{2}}{0.11} \right)^{\frac{2}{3}} \text{cm}. 
\end{equation}

\noindent For M15 X-3, this gives $a \sim 1.3 \times 10^{11}$ cm. This means that the X-ray flux received by the companion at its surface \refbf{(assuming isotropic emission)} is

\begin{equation}
F_{x} \approx \frac{L_{x}}{4 \pi a^{2}} = 3.5 \times 10^{10} \text{ erg cm$^{-2}$ s$^{-1}$}.
\end{equation}

By contrast, the total flux of X-3's companion can be estimated by taking a luminosity of $L_{\bigstar} \approx 0.05 L_{\sun}$ \citep{Baraffe97} and $R_{\bigstar} \approx 0.44 R_{\sun}$ \citep{FKR02} to get the companion's flux at its surface to be:

\begin{equation}
F_{\bigstar} \approx \frac{L_{\bigstar}}{4 \pi R_{\bigstar}^{2}} = 1.6 \times 10^{10} \text{ erg cm$^{-2}$ s$^{-1}$}.
\end{equation}

Since the received X-ray flux is roughly a factor of 2 larger than the flux emitted by the companion itself, X-ray irradiation is likely to be a contributing factor to the optical color of this system. This means that it is likely that the companion of M15 X-3 is less massive than the fitted mass of $0.440^{+0.035}_{-0.060}$ \Msun. To constrain the mass of the companion, we assumed that the companion is irradiated by \refbf{its primary with $L_{\rm primary} = 10^{34}$ erg s$^{-1}$} and re-radiates these X-rays away as a blackbody of temperature
\begin{equation}
T_{irradiation} = \left(\frac{\sigma}{F_{\bigstar} + F_{x}}\right)^{\frac{1}{4}}
\end{equation}

We assume that the emission is dominated by the irradiated half of the star, and thus that:

\begin{equation}
\left(\frac{R}{R_{\sun}}\right)^{2} = \left(\frac{L}{L_{\sun}}\right) \left(\frac{T}{T_{\sun}}\right)^{-4}
\end{equation}

\begin{figure}
\includegraphics[scale=0.4]{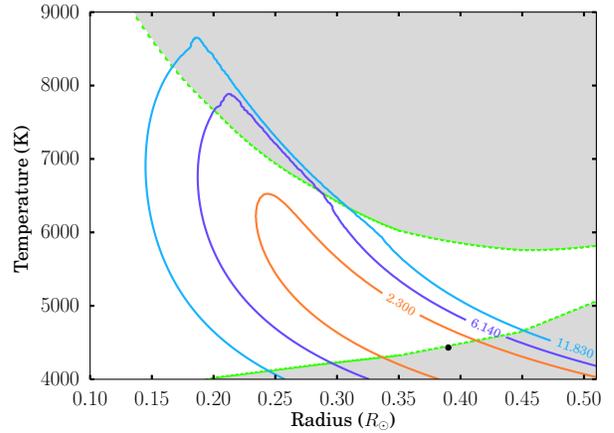}
\caption{%
Contour plot of the fit of a single power law and irradiated companion (represented as a blackbody) for varying choices of the companion size and temperature. The orange, purple, and cyan curves indicate the $1\sigma$, $2\sigma$, and $3\sigma$ curves, respectively. The bottom green curve indicates the temperature/radius relation for non-irradiated low-mass main-sequence stars of metallicity $[M/H] = -2.0$, while the top green curve represents the temperature/radius relation for the same stars maximally irradiated by $10^{34}$ erg s$^{-1}$ of X-rays. \refbf{For reference, the black point indicates the best-fit $0.44 M_{\sun}$ star for the non-irradiated case. Regions in grey are excluded for physical reasons - those below the non-irradiated main sequence are aphysically colder than a main-sequence star of the same radius, while those above the maximally irradiated main sequence are hotter than possible for a main-sequence star irradiated by only a primary with an X-ray luminosity of $10^{34}$ erg s$^{-1}$.}
\label{irradiation} 
}
\end{figure}

Modelling the star as a blackbody with temperature $T$ and radius $R$, the acceptability of various fits was explored over the parameter space in Figure \ref{irradiation}. When compared with a maximally irradiated main sequence star, at the $3\sigma$ level we can still rule out the possibility of a $0.15 R_{\sun}$ or smaller companion. This rules out a brown dwarf, white dwarf, or planetary companion. At the $2\sigma$ level, the radius of the companion lies between 0.19 and 0.41 $R_{\sun}$, which suggests a $\sim 2-4$ hour orbital period.

\section{Discussion}\label{s:discuss}

\begin{figure}
\includegraphics[scale=0.45]{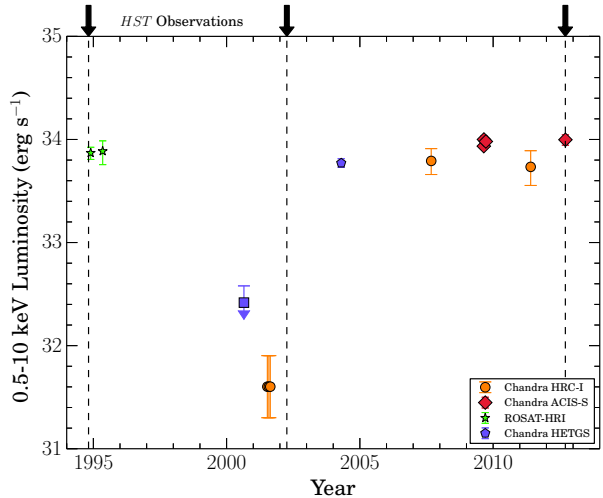}
\caption{%
Long term light curve for M15 X-3. ROSAT, \Chandra\ HETGS, and \Chandra\ HRC-I data (where M15 X-3 is in quiescence) are taken from \cite{Heinke09}, while the \Chandra\ HRC-I (where M15 X-3 is bright) and ACIS-S data were analyzed in this paper. The vertical dashed lines indicate HST observing epochs, including the 2012 observation that is nearly simultaneous with an ACIS-S observation.
\label{longterm}
}
\end{figure}

\subsection{Nature of System}
M15 X-3's unusual X-ray behavior for a globular cluster X-ray source, and position blue-ward of the main sequence, suggests that we check whether it can be explained as a foreground or background source. If M15 X-3 were a foreground star, we would expect it to be redder, rather than bluer, than the main sequence. We use the empirical relations from \cite{Baldi02} to estimate the probability of finding a background AGN at or above the flux of M15 X-3 projected within the half-mass radius of M15. For M15 X-3, we use the bright state flux of $S \approx 5.6 \times 10^{-13}$ erg/(s cm$^{2}$), which gives an expected 0.6 sources per square degree, or $5 \times 10^{-4}$ sources expected in the half-mass radius of M15, roughly the region where M15 X-3 was discovered. (Note that the likelihood of finding a source at the distance of M15 X-3 from the cluster center is a factor of 9 smaller.) This suggests that M15 X-3 is unlikely to be a foreground or background source. 

\refbf{We can quantitatively consider the probability of the optical counterpart being an unaffiliated interloper by considering the local density of optical sources that could be potential counterparts. We consider a region of $5"$ around M15 X-3, as there is a strong density gradient of sources with distance from the center of the cluster. Within this region, there are 3 objects that have V magnitudes in the range $19 < V < 23 $ that lie as far to the blue side of M15's main sequence as M15 X-3's proposed counterpart, including the counterpart itself. This gives a chance probability of one of these objects falling within the M15 X-3 error circle (0.05" --- the offset between the \Chandra\ and \HST\ relative astrometry, determined using the optical and X-ray detections of the nearby XRB AC 211 \citealt{Heinke09}) of $3*(0.05"/5")^{2} \approx 3 \times 10^{-4}$. Standard counting errors would give $3^{+2.9}_{-1.6}$ objects at $1\sigma$, and $3^{+9.7}_{-2.8}$ at $3\sigma$. At the upper end, the probability of a non-associated optical counterpart that is a member of the cluster is $\approx 0.13\%$ even at $3\sigma$. The remaining possibility for the optical emission of M15 X-3 is that the reddish companion is a cluster member unassociated with M15 X-3 itself. However, if the B flux is dominated by M15 X-3, and the I flux is dominated by an unassociated main sequence star, the offset between these two objects could not be more than about 1 WFC3 pixel = 0.04", as the counterpart is well-aligned in the B and I frames. This gives a chance probability of an unassociated red star happening to align with M15 X-3's blue counterpart of $\approx 3\%$, considering stars within one magnitude of the proposed counterpart in I.}

As discussed in \cite{Heinke09}, we can immediately rule out a white dwarf accretor on the basis of long-term variability and the X-ray to optical flux ratio. Radio observations of M15 X-3 argue against the idea that we view it edge-on, permitting most of its intrinsic luminosity to be obscured by the accretion disk, because the radio emission, which scales with the intrinsic X-ray luminosity, should still be visible (e.g., \citealt{Bower05}). \cite{Strader12} report no source at M15 X-3's location, with an rms sensitivity of 2.1 $\mu$Jy/beam. A non-detection implies that M15 X-3's radio brightness cannot be higher than 10.5 $\mu$Jy at 5$\sigma$. For a black hole at $10^{36}$ erg s$^{-1}$, with X-rays obscured by a high inclination accretion disk, we would expect a radio flux of roughly $10^{30}$ erg s$^{-1}$. At 10.3 kpc, this would give roughly 1 mJy at 6 GHz \citep{Gallo2012}. For a neutron star, we would expect closer to 20 $\mu$Jy at 6 GHz \citep{Migliari06}. 
\subsection{X-ray Spectra}
M15 X-3 appears to be the faintest VFXT for which we have an X-ray spectrum of very high quality; for instance, the targets discussed in \cite{ArmasPadilla13} have $L_X$(0.5-10 keV) of 5-10$\times10^{34}$ erg s$^{-1}$, vs. $\sim1\times10^{34}$ erg s$^{-1}$ here.  The X-ray spectrum is somewhat atypical, and can be well-fit by either a broken power-law  or a power-law plus a thermal component.  The thermal component, if fit by a blackbody, produces an inferred radius of 0.2$^{+0.1}_{-0.1}$ km (or 0.3$^{+0.4}_{-0.2}$ km if fit with an NSA model), which is significantly smaller than the inferred blackbody radii ($\sim$5 km) of the thermal components in the persistent VFXTs discussed by \cite{ArmasPadilla13}. It is unclear if the possible thermal component here has the same physical origin as in the other transients.  

The X-ray spectrum of material falling onto a neutron star is predicted to include a thermal component (from the lower atmosphere) plus a very hot surface layer that effectively Comptonizes the outgoing radiation \citep{Deufel01}. This model may provide a first step toward understanding the spectra of several observed neutron star transients at low luminosity (e.g., \citealt{ArmasPadilla13}, \citealt{Bahramian14}, \citealt{Chakrabarty14}).  In the case of M15 X-3, the total luminosity and small inferred radius of the thermal component are consistent with such a high proton energy flux ($\sim 10^{24} {\rm \, erg \, cm^{-2} \, s^{-1}}$) that the \cite{Deufel01} model predicts an entirely non-thermal spectrum. To reproduce the approximately 10\% thermal component in M15 X-3, one might invoke a central high proton flux hotspot that is smaller than the radius of the thermal component we measure, where the thermal component arises from the surrounding area being heated either by conduction or by a lower proton energy flux. It would be of great interest to obtain models such as Deufel's in a convenient form for the spectral fitting of X-ray transients \footnote{The {\it zamp} models of \cite{Zampieri95} do not include the hot Comptonizing layer of the atmosphere.}.

\subsection{Explanation for VFXT Behavior}

If M15 X-3 spends most of its time in its "bright" state, with a persistent luminosity of $10^{34}$ erg s$^{-1}$, then we can infer an upper limit on the time-averaged accretion rate by assuming that all of its 0.5-10 keV X-ray luminosity is from accretion. This assumption implies that M15 X-3's time-averaged accretion rate could be at most $9\times10^{-13}$ \Msun/year. Under conventional binary evolution, the low X-ray luminosity of M15 X-3 is difficult to explain. One suggestion is that the low accretion rates needed for VFXTs is possible from a brown dwarf or planetary-mass companion. Another option is a system where the accretor is an intermediate-mass ($\sim$1000 \Msun) black hole with a primordial companion that is very low mass in the current epoch \citep{KingWijnands06}. If the companion of M15 X-3 is a brown dwarf or other very low-mass companion, then we would expect the optical emission from the system to be produced only by the accretion disk. At 10.3 kpc, a brown dwarf companion would be well below the detection threshold from the \HST\ observations, in contrast with the presence of a second, non-accretion disk component. The poor fit of a power law alone to the BVI values for M15 X-3 suggests that a disk-only model is insufficient to explain the nature of M15 X-3 --- the accretor's companion is not sufficiently low-mass for this explanation to be tenable. The location of M15 X-3, 21" from the cluster center, places it well outside the region where an intermediate-mass black hole would be expected to wander \citep{Chatterjee02,Gerssen02}. 

It is also plausible for some VFXTs that accretion from the wind of a MS star could explain low accretion rates. In the case of M15 X-3, a companion of $ M \lesssim 0.44 M_{\sun} $ is unlikely to be driving wind-fed accretion in this system \citep{Heinke14}. It has also been argued by \citet{intZand07} that persistently low accretion rates require a very small, completely ionized disk, implying an ultracompact system. Our optical observations of M15 X-3 are inconsistent with an ultracompact system, as we have found evidence for a relatively large (0.15-0.45 $R_{\odot}$) donor.

An alternative explanation for the unusually low X-ray persistent X-ray luminosity of M15 X-3 could be that it is an otherwise normal LMXB (with a 2-4 hour orbital period) where the NS magnetic field is sufficiently strong to interfere with accretion, via the propeller effect (in which the rotation of the NS's magnetic field is able to repel inflowing matter; \citealt{Illarionov75}). \cite{DAngelo12} modelled a leaky propeller and "trapped" disk, leading to continuous low-level accretion; similarly leaky propellers have been studied by e.g. \cite{Romanova03}, \cite{Ustyugova06}, \cite{Kulkarni08}. \cite{Heinke14} argued that all the quasi-persistent VFXTs such as M15 X-3 are in this physical state. \refbf{This argument is based, on one hand, on the difficulty of explaining continued extremely low-level accretion via disk instability models, and on the other hand the example of transitional millisecond pulsars, which maintain low-level accretion and clearly show evidence of magnetic effects on the accretion flow.}

Three transitional millisecond pulsars (which cycle between states of low-level accretion at $10^{33}-10^{34}$ erg s$^{-1}$, non-accreting radio pulsar states at $\sim10^{31}$ erg s$^{-1}$, and occasional higher-$L_X$ accreting states) have recently been identified \citep{Archibald09,Papitto13b,Patruno14,Stappers14,Bassa14,Bogdanov14,Roy14}. The recent detection of X-ray pulsations from two of these transitional millisecond pulsars in their $L_X \sim 10^{33}-10^{34}$ erg s$^{-1}$ state \citep{Archibald14,Papitto14} strongly indicates that these pulsars are indeed in this leaky propeller state \citep{DAngelo14}. The quasi-persistent VFXTs, including M15 X-3, may therefore also be transitional millisecond pulsars \citep{Degenaar14,Heinke14}. \refbf{At least one transitional millisecond pulsar has been observed to be relatively radio-bright in its X-ray active phase \citep{Deller14}. Of the known tMSP systems, M15 X-3's X-ray luminosity is closest to XSS J12270-4859, which would imply it has a radio brightness of roughly $5 \times 10^{27}$ erg s$^{-1}$ at 5 GHz if it is a similar source. Assuming a spectrally flat source, M15 X-3 would have a 5 GHz flux density of approximately 5.3 $\mu$Jy at 10.3 kpc, roughly a factor of two too faint to be detected by existing radio observations.}

\refbf{A relatively conservative estimate of M15 X-3's duty cycle is roughly $70\%$, as seen in Figure \ref{longterm}. However, the quasi-persistent VFXTs with a duty cycle resembling that of M15 X-3 tend to have relatively soft power-law fits, or harder power-law fits only when a thermal component is also present \citep{ArmasPadilla13}}. The X-ray spectrum of M15 X-3 is unlikely to be an intrinsically harder spectrum being suppressed at lower energies because of absorption effects --- radio observations suggest the system is not being viewed edge-on, and \refbf{the column density of M15 X-3 we measure is only $\lesssim 10^{21} \rm cm^{-2}$.}

\section{Conclusion}

We have performed a combined optical/X-ray followup observation of the VFXT M15 X-3. Our X-ray observation, and other archival \Chandra\ observations, found M15 X-3 in its "bright" state, persistently at a luminosity of $\approx 1 \times 10^{34}$ erg s$^{-1}$. The X-ray spectrum, if fit with a power-law, show strong residuals. The X-ray spectra are better fit by a power law of index $1.28^{+0.06}_{-0.06}$ up to an apparent spectral break at $2.7^{+0.3}_{-0.6}$ keV, where it is described by a power law of index $1.9^{+0.2}_{-0.2}$. A power-law plus thermal component also describes the spectrum acceptably, giving an emitting region of $\sim 0.21$ km for  $kT \sim 0.6$ keV (for a blackbody model) or $\sim 0.4$ km for $kT \sim 0.6$ keV (for a NS atmosphere model). The \HST\ observation clearly detects a blue optical counterpart, which includes contributions from both the system's accretion disk and a companion with emission consistent with a main-sequence star of mass $0.440^{+0.035}_{-0.060}$ \Msun. The actual mass may be lower, in the range 0.20-0.45 \Msun (at $2\sigma$), since X-ray irradiation of the companion can cause the star to emit more strongly in the optical bands than it would as an isolated main-sequence star. M15 X-3's persistently low accretion rate is difficult to explain in conventional binary evolution models. Combining this low accretion rate, with the evidence of the donor nature presented here, argues for accretion via a leaky propeller mechanism, as has been recently suggested for the transitional radio pulsars. 

\section{Acknowledgments}

This research made use of the NASA Astrophysics Data System (ADS), as well as the \textit{CIAO} software package developed by the Chandra X-Ray Center (CXC) and the \textit{pysynphot} package developed as part of STSDAS by the Space Telescope Science Institute (STScI). R.M.A. is supported by a Queen Elizabeth II Scholarship. R.M.A., G.R.S., and C.O.H.~are supported by NSERC Discovery Grants. C.O.H. is also supported by an Alberta Ingenuity New Faculty Award and an Alexander von Humboldt Fellowship. H.C. and P.L. acknowledge the support of NASA Chandra grant GO2-13046X and NASA HST grant GO-12751.03 to Indiana University.  We made extensive use of observations from the Chandra X-Ray Observatory (CXO), Chandra Data Archive, and the Hubble Space Telescope (HST). We thank the referee for helpful comments during the review process. We also thank D. Grant for helpful discussion. 

\bibliographystyle{apj}
\bibliography{references}

\end{document}